\documentclass[sigconf, table]{acmart}
\usepackage{xcolor}
\usepackage{adjustbox}
\usepackage{graphicx}
\graphicspath{{./figs/}}
\usepackage{balance}
\usepackage{xspace}
\usepackage{url}
\usepackage{booktabs}

\usepackage{enumitem}
\usepackage{multirow}
\usepackage{amsmath}
\usepackage{amssymb}

\usepackage{subfigure}
\usepackage{tikz}
\usepackage{wasysym}
\usepackage{textcomp}
\newcommand{\rmnum}[1]{\romannumeral #1}
\newcommand{\Rmnum}[1]{\expandafter\@slowromancap\romannumeral #1@}

\usepackage{algorithmic}
\usepackage{hyperref}

\usepackage[ruled,lined,linesnumbered,vlined,algo2e]{algorithm2e}
\usepackage{url}

\usepackage{natbib}
\usepackage{pifont}

\newcommand{\etal}{\hbox{{et al.}}\xspace}
\newcommand{\eg}{\hbox{{e.g.}}\xspace}
\newcommand{\ie}{\hbox{{i.e.}}\xspace}
\newcommand{\wrt}{\hbox{{w.r.t.}}\xspace}
\newcommand{\etc}{\hbox{{etc.}}\xspace}

\newcommand{\aka}{\hbox{{a.k.a.}}\xspace}

\newcommand{\ausera}{{\textsc{Ausera}}\xspace}

\newcommand{\Weakness}{{Weakness}\xspace}
\newcommand{\Weaknesses}{{Weaknesses}\xspace}
\newcommand{\weakness}{{weakness}\xspace}
\newcommand{\weaknesses}{{weaknesses}\xspace}

\newcommand{\app}{{banking app}\xspace}
\newcommand{\apps}{{banking apps}\xspace}

\newcommand{\Apps}{{Banking apps}\xspace}

\newcommand{\feedback}{21\xspace}
\newcommand{\patched}{16\xspace}
\newcommand{\all}{2,157\xspace}

\SetAlFnt{\small}
\SetAlCapFnt{\small}
\SetAlCapNameFnt{\small}
\SetAlCapHSkip{0pt}
\IncMargin{-\parindent}
\definecolor{pblue}{rgb}{0.13,0.13,1}
\definecolor{pgreen}{rgb}{0,0.5,0}
\definecolor{pred}{rgb}{0.9,0,0}
\definecolor{pgrey}{rgb}{0.46,0.45,0.48}
\usepackage{listings}
\def\WithComments{}
\ifdefined \WithComments

\newcommand{\sen}[1]{\textcolor{black}{#1}}

\newcommand{\revised}[1]{\textcolor{black}{#1}}
\else

\newcommand{\sen}[1]{\textcolor{black}{#1}}

\fi

\copyrightyear{2020}
\acmYear{2020}
\setcopyright{acmcopyright}
\acmConference[ICSE '20]{42nd International Conference on Software
	Engineering}{May 23--29, 2020}{Seoul, Republic of Korea}
\acmBooktitle{42nd International Conference on Software Engineering (ICSE '20), May
	23--29, 2020, Seoul, Republic of Korea}
\acmPrice{15.00}
\acmDOI{10.1145/3377811.3380417}
\acmISBN{978-1-4503-7121-6/20/05}

\begin{document}
\title{An Empirical Assessment of Security Risks of Global Android Banking Apps}
	
\author{Sen Chen$^{1}$, Lingling Fan$^{1}$, Guozhu Meng$^{2}$$^,$$^{3}$, Ting Su$^{4}$, Minhui Xue$^{5}$, Yinxing Xue$^{6}$} 
\author{Yang Liu$^{1}$$^,$$^{8}$, Lihua Xu$^{7}$}
\renewcommand{\authors}{Sen Chen, Lingling Fan, Guozhu Meng, Ting Su, Minhui Xue, Yinxing Xue, Yang Liu, Lihua Xu}
\affiliation{$^{1}$Nanyang Technological University, Singapore}
\affiliation{$^{2}$SKLOIS,  Institute of Information Engineering,  Chinese Academy of Sciences,  China}
\affiliation{$^{3}$School of Cyber Security, University of Chinese Academy of Sciences, China $^{4}$ETH Zurich, Switzerland}
\affiliation{$^{5}$The University of Adelaide, Australia $^{6}$University of Science and Technology of China, China}
\affiliation{$^{7}$New York University Shanghai, China $^{8}$Zhejiang Sci-Tech University, China}
\email{chensen@ntu.edu.sg}
	
\renewcommand{\shortauthors}{S.~Chen, L.~Fan, G.~Meng, T.~Su, M.~Xue, Y.~Xue, Y.~Liu, and L.~Xu}


	
\begin{abstract}
	Mobile banking apps, belonging to the most security-critical app category, render massive and dynamic transactions susceptible to security risks. Given huge \revised{potential} financial loss caused by vulnerabilities, existing research lacks a comprehensive empirical study on the security risks of global banking apps to provide useful insights and improve the security of banking apps.
	
	Since data-related weaknesses in banking apps are critical and may directly cause serious financial loss, this paper first revisits the state-of-the-art available tools and finds that they have limited capability in identifying data-related security weaknesses of banking apps. To complement the capability of existing tools in data-related weakness detection, we propose a three-phase automated security risk assessment system, named \ausera, which leverages static program analysis techniques and sensitive keyword identification. By leveraging \ausera, we collect 2,157 weaknesses in 693 real-world banking apps across 83 countries, which we use as a basis to conduct a comprehensive empirical study from different aspects, such as global distribution and weakness evolution during version updates. We find that apps owned by subsidiary banks are always less secure than or equivalent to those owned by parent banks. In addition, we also track the 
	\revised{patching of weaknesses} and receive much 
	\revised{positive} feedback from banking entities so as to improve the security of banking apps in practice. We further find that weaknesses derived from outdated versions of banking apps or third-party libraries are highly prone to being exploited by attackers. To date, we highlight that 21 banks have confirmed the weaknesses we reported (including 126 weaknesses in total). We also exchange insights with 7 banks, such as HSBC in UK and OCBC in Singapore, via in-person or online meetings to help them improve their apps. We hope that the insights developed in this paper will inform the communities about the gaps among multiple stakeholders, including banks, academic researchers, and third-party security companies. 
\end{abstract}

\maketitle

\section{Introduction}\label{sec:introduction}
Banking apps belong to the most security-critical and data-sensitive app category. Cashless mobile payment has significantly fragmented the traditional financial services, beginning with the first ATM and culminating in e-banking. 
Users often misconceive that \apps provide secure transactions and an easy-to-use interface, by assuming all communications are done between local \apps and remote bank servers securely (e.g., over HTTPS).
Unfortunately, this assumption does not always hold. 
After examining many real-world banking apps, we find new types of weaknesses that 
\revised{are hard to be detected} 
by existing industrial and open-source tools, {e.g., QiHoo360~\cite{tool_360}, AndroBugs~\cite{androbugs}, MobSF~\cite{mobsf}, and QARK~\cite{qark}.} 
For example, in a popular banking app from Google Play, the user will be asked to register with her personal information, including first name, last name, password, and address. 
After the user clicks the ``register'' button, the app sends an SMS attached with the sensitive data (in plain text) to authenticate that user, but the data is stored in the SMS outbox unexpectedly. 
If an attacker registers a content observer to the SMS outbox on the mobile device with \texttt{READ\_SMS} permission, the user's sensitive data can be easily intercepted by the attacker.
Indeed, many other real-world banking-specific weaknesses and attacks have been witnessed globally
~\cite{appknox2015,appknox2016}.
Nowadays, banking apps pose new challenges, such as flaws and vulnerabilities~\cite{attack1, attack3} that cause huge financial loss~\cite{attack2}.

\begin{figure*}
	\centering	\includegraphics[width=0.925\textwidth]{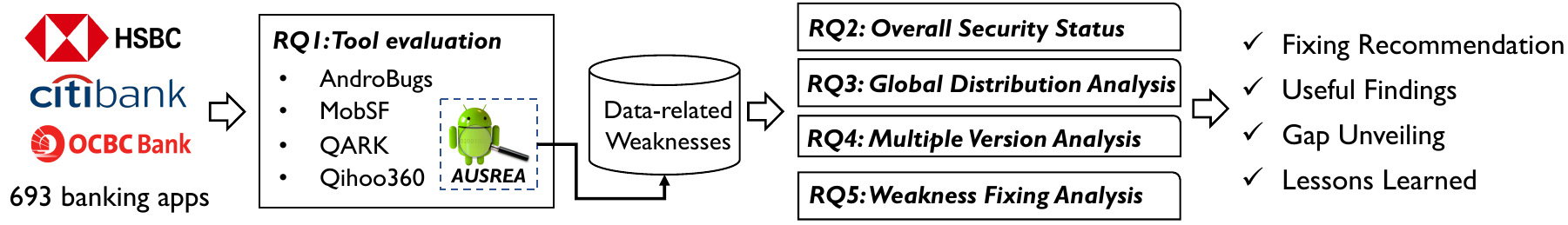}
	\caption{Overview of our study}
	\label{fig:overview}
\end{figure*}

To understand the weaknesses exhibited in banking apps and help to improve the security of these apps, {several studies have} been done manually on a small-scale banking apps~\cite{reaves2015mo,castle2016let,parasa2016mobile, reaves2017mo}. 
The conclusions drawn from manual analysis may be more likely to be biased and cannot represent the security status of the entire banking ecosystem.
Apart from manual analysis on only a small-scale apps, state-of-the-art assessment approaches also pose several other limitations:
(1) current studies lack a baseline
of sensitive data-related security \weaknesses specific to the core functionality of banking apps to ensure an overall assessment of these apps;
(2) the current off-the-shelf services (\eg, \textsc{Qihoo360}~\cite{tool_360}) and open-source tools (\eg, \textsc{AndroBugs}~\cite{androbugs}) use syntax-based scanning to perform a security check during app development, which would incur a large number of false positives (\eg, non-sensitive data printed in the log file). Besides, these tools focus on generic categories of apps, not specific to banking apps. 
Even when the weaknesses, such as cryptographic misuses~\cite{egele2013empirical} and inappropriate SSL/TLS implementations~\cite{fahl2012eve,georgiev2012most,sounthiraraj2014smv,chothia2017banker}, have been reported for years, it still appears unknown why so many security weaknesses in \apps are not yet patched~\cite{reaves2017mo}.
{Overall, the existing work cannot represent the security status of the entire banking ecosystem, and the state-of-the-art tools are ineffective in collecting a large number of weaknesses to conduct further in-depth analysis.}

To explore the entire mobile banking ecosystem and help to ensure the user's financial security, this paper takes a large number of \apps as subjects to conduct a comprehensive empirical study on the data-related weaknesses in global Android banking apps.
As shown in Figure~\ref{fig:overview}, 
our study contains three main steps: 
(1) we first collect 693 banking apps across 83 countries from various markets, to our knowledge, this is the largest banking app dataset taken into study to date;
(2) to collect the weaknesses exhibited in banking apps and complement the capability of existing tools in data-related weakness detection, we first summarize a weakness baseline and propose an \underline{\bf au}tomated \underline{\bf se}curity \underline{\bf r}isk \underline{\bf a}ssessment system (\ausera). \ausera combines static program analysis techniques and sensitive keyword identification, to identify such weaknesses ({cf. Section~\ref{sec:tool_evaluation}}).
(3) By applying \ausera, we collected 2,157 security weaknesses in the 693 banking apps, and further conduct a comprehensive empirical study ({cf. Section~\ref{sec:study}}) to investigate the ecosystem of banking apps in terms of security weaknesses, aiming to answer the following research questions:

\begin{itemize}
	
	\item \textbf{RQ1:} What is the current status of existing tools towards collecting reliable data-related weaknesses in banking apps compared with \ausera?
	
	\item \textbf{RQ2:} What is the overall security status of banking apps in terms of data-related weaknesses?
	
	\item \textbf{RQ3:} What is the weakness status of \apps globally \wrt economies and regulations? 
	
	\item \textbf{RQ4:} How are weaknesses introduced during app evolution and fragmentation?
	
	\item \textbf{RQ5:} What is the gap between academic researchers and banks 
	\revised{in understanding and fixing weaknesses?}
\end{itemize}

{Through an in-depth analysis of the weaknesses,}
we find that (1) banking apps across different regions exhibit various types of security status, mainly due to different economy status (\eg, small village banks) and financial regulations (\eg, {GDPR}~\cite{gdpr}).
Banking apps in Europe and North America have few security \weaknesses, with only 0.27 weakness of data leakage per app. Asia is most flooded with security weaknesses, averaging out to 6.4 \weaknesses per app. Banking apps from Africa have comparatively moderate security status with 4.6 \weaknesses per app, primarily because of its high demand for cashless payment services. 
(2) \Weaknesses of apps vary across different markets by countries and bring fragmentation problems among different versions of the same banking {apps}. 
Apps owned by subsidiary banks are always less secure than or equivalent to those owned by parent banks. 
This observation is evidenced by the South Korean version of the Citibank app and the Chinese version of the HSBC app.
(3) Apart from the lessons learned from our study, we also track the weakness fixing process based on our reported weaknesses and set up 9 in-person or online meetings with 7 banks. 
These meetings help the communities understand the gaps between different parties, including banks, academic researchers, and third-party security companies.

In summary, we make the following contributions:

\begin{itemize}
	
	\item To collect weaknesses in banking apps and complement the capability of existing tools in data-related weakness detection, we developed an automated security risk assessment system (\ausera), to efficiently identify security \weaknesses in banking apps, outperforming 4 state-of-the-art industrial and open-source tools.
	
	\item To our knowledge, we conducted the first large-scale empirical study on 2,157 security weaknesses collected from 693 banking apps, the largest dataset taken into study to date. 
	We attempt to investigate the ecosystem of global banking apps in terms of data-related weaknesses from four different aspects, such as global distribution analysis and evolution of multiple versions.
	
	\item We report the identified weaknesses to banks and provide simple-but-concrete fixing recommendations. To date, \feedback banks have acknowledged our results, and 52 reported \weaknesses have been patched by the corresponding banks. Some of them have actively collaborated with us to improve the security of their apps.
\end{itemize}

\section{Tool Evaluation}\label{sec:tool_evaluation}
In this section, we propose an automated weakness detection tool (named \ausera), guided by our constructed security weakness baseline in order to collect security weaknesses in banking apps. We also evaluate its effectiveness compared with the state-of-the-practice tools to observe the current status of detection ability towards data-related weaknesses in banking apps.
We then introduce the data collection process of banking apps and security weaknesses in these apps as the basis to conduct a large-scale analysis.
{Before proposing \ausera, we first revisit the state-of-the-art available tools or online services for weakness detection.}

\textsc{AndroBugs}~\cite{androbugs}, \textsc{QARK}~\cite{qark}, and \textsc{MobSF}~\cite{mobsf} are all open-source tools for detecting vulnerabilities in general Android apps. Specifically, 	
\textsc{AndroBugs} is a framework to find potential vulnerabilities in Android apps by pattern-matching, and it also records some meta data in the database such as permissions used in the current app.
\textsc{QARK} is designed to look for vulnerabilities related to Android apps, either in source code or packaged APKs.
\textsc{MobSF} is a pen-testing framework, which is able to detect app vulnerabilities, and the results can be displayed on webpages.
Apart from the open-source detection tools, \textsc{Qihoo360} is a popular security company in China, which maintains an app scan engine, named \textsc{appscan}~\cite{tool_360}. It is a free online application for security risk scanning service.

However, the current off-the-shelf services and tools have the following limitations in banking specific weakness collection {according to our investigation}: 
(1) They usually use syntax-based scanning, thus cannot verify the actual data flow, which would incur a large number of false positives that are not related to sensitive data leakage. 
(2) They usually aim to detect weaknesses in general apps, not specific to banking apps. Thus the patterns they use to detect weakness are difficult to detect 
{data-related} weaknesses {in banking apps}.
The detection ability of the state-of-the-art weakness detection tools are demonstrated in Section~\ref{sec:evaluation}.
Considering the aforementioned situations, to complement the capability of existing tools in data-related weakness detection, we propose a tool, \ausera, for automating the detection and collection of sensitive-date related weaknesses specific to banking apps.

\begin{table}\footnotesize
	\renewcommand{\arraystretch}{1.2}
	\centering
	\caption{Taxonomy of security weaknesses in our study}
	\label{tbl:issues}
	\begin{adjustbox}{max width=\textwidth}
		\begin{tabular}{l|l|l}
			\hline
			{\begin{tabular}[c]{@{}l@{}} {\bf Category}\end{tabular}}
			& \multicolumn{2}{l}{\begin{tabular}[c]{@{}l@{}} {\bf  Security \Weakness Type} \end{tabular}}
			\\ \hline
			
			{\multirow{1}{*}{\begin{tabular}[c]{@{}l@{}}{\bf Input Harvest}\end{tabular}}}
			& \multicolumn{2}{l}{Sensitive data harvested by screenshots} \\ \hline
			
			{\multirow{5}{*}{\begin{tabular}[c]{@{}c@{}}{\bf Sensitive}\\ {\bf Data Storage}\end{tabular}}}
			
			& \multicolumn{2}{l}{Stored in shared preferences} \\ 
			
			&  \multicolumn{2}{l}{Stored in webview.db} \\ 
			
			& \multicolumn{2}{l}{Logged locally}  \\ 
			
			& \multicolumn{2}{l}{Stored on SD Card} \\ 
			
			& \multicolumn{2}{l}{Written in text files}  \\ \hline
			
			{\multirow{4}{*}{\begin{tabular}[c]{@{}c@{}}{\bf Sensitive Data} \\{\bf Transmission}\end{tabular}}}
			
			& \multicolumn{2}{l}{Transmitted via {SMS}}  \\ \cline{2-3}
			
			& \multirow{3}{*}{\begin{tabular}[c]{@{}l@{}}ICC leaked\end{tabular}}
			&by dynamically {registered} {Receiver}  \\ 
			
			&  & by implicit {Intent} \\ 
			
			&  &by {component} export \\ \hline
			
			{\multirow{12}{*}{\begin{tabular}[c]{@{}l@{}}{\bf Communication} \\ ~~~~~~{\bf Infrastructure}\end{tabular}}}
			
			& \multicolumn{2}{l}{Only uses {HTTP} protocol} \\ 
			
			& \multicolumn{2}{l}{Uses invalid certificates (\ie, expiration, SHA-1 used)}  \\ \cline{2-3}
			
			& \multirow{3}{*}{\begin{tabular}[c]{@{}l@{}}Uses invalid \\  certificate \\ authentication\end{tabular}} & allows all hostname request   \\ 
			
			&  & uses invalid hostname verification   \\ 
			
			&  & uses invalid server verification  \\ \cline{2-3}
			
			& \multicolumn{2}{l}{Uses hard-coded encryption key} \\ \cline{2-3}
			
			& \multirow{2}{*}{\begin{tabular}[c]{@{}l@{}}Uses improper \\ AES encryption\end{tabular}}
			
			&  uses insecure encryption \\ 
			
			&  & uses improper function\\ \cline{2-3}
			
			& \multirow{2}{*}{\begin{tabular}[c]{@{}l@{}}Uses improper \\ {RSA} encryption\end{tabular}}
			
			&  no {RSA} \\ 
			
			&  &uses improper function \\  \cline{2-3}
			
			& \multicolumn{2}{l}{\begin{tabular}[c]{@{}l@{}} Uses insecure SecureRandom (\ie, setSeed) \end{tabular}}  \\ 
			
			& \multicolumn{2}{l}{Uses insecure hash function (\ie, {MD5} and {SHA-1})} \\
			\hline
			
		\end{tabular}
	\end{adjustbox}
\end{table}

\subsection{\ausera}
In order to collect a data-related weakness dataset specific to banking apps, we first propose a taxonomy of sensitive  data-related security weaknesses in banking apps. Guided by the baseline, \ausera is proposed to identify weaknesses in banking apps. 
\subsubsection{\textbf{Taxonomy of Security \Weaknesses within Banking Apps}}\label{sec:baseline}

\revised{We propose and integrate security \weaknesses of mobile banking apps from prior research~\cite{reaves2015mo,castle2016let,parasa2016mobile,reaves2017mo}},
best industrial practice guidelines and reports (\eg, OWASP~\cite{owasp_mobile_security}, Google Android Documentation~\cite{google_best_practice}, and AppKnox security reports~\cite{appknox2015,appknox2016}), NowSecure reports~\cite{nowsecure2017}, and security weakness and vulnerability databases (\eg, CWE~\cite{cwe}, CVE~\cite{cve}).
We take an in-depth look at   
the \weaknesses \wrt \textbf{\emph{sensitive data}}, since the biggest threat to \apps comes from manipulation of digital assets and routine financial activities. 
As shown in Table~\ref{tbl:issues}, sensitive data may be exposed to attackers {through various ways as follows:} 

\begin{itemize}[noitemsep,topsep=0pt,leftmargin=*]
	\item \textit{Input Harvest,} confidential inputs and user relevant sensitive data (\eg, transaction details) can be harvested via UI screenshot by malicious apps on rooted devices, or even adb-enabled devices without root access~\cite{lin2014screenmilker}.
	\item \textit{Data Storage,} an adversary is able to obtain data stored in local storage (\eg, shared preference, webview.db) on rooted devices or external storage (\eg, SD Card), \sen{and also from the output of the Android logging system without root access}.
	\item \textit{Data Transmission,} sensitive data transmission via {SMS} can be easily intercepted by malware observing the outbox of Android {SMS} service. 
	Moreover, data leakage via inter-component communication (ICC) is another {potential} threat, allowing third parties to obtain data from \apps by making implicit intent calls, or dynamic registration of a broadcast Receiver.
	\item \textit{Communication Infrastructure,} MITM attack can obtain sensitive data through sniffing network traffic between client and server, thereby sending fake data to either party. This kind of attack is generally achieved due to improper authentication protocols, insecure cryptography, lack of certificate verification, \etc
\end{itemize}

\revised{Our baseline contains data-related \weaknesses of multiple categories and builds a solid foundation for analyzing \weaknesses 
{in} banking apps.}

\begin{figure}[t]
	\centering
	\includegraphics[width=0.5\textwidth]{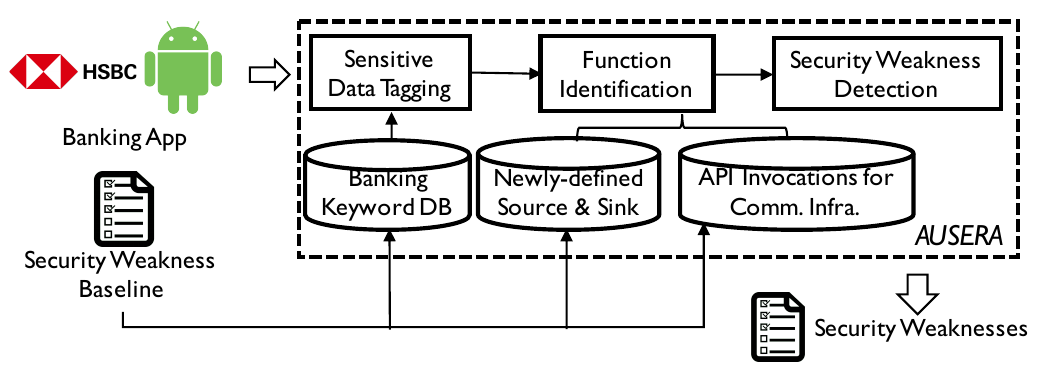}
	\caption{\sen{Overview of \ausera}}
	\label{fig:architecture}
\end{figure}

\subsubsection{\textbf{Methodology of \ausera}}
To collect a large dataset of security weaknesses,
\sen{\ausera takes as input each banking app, guided by the weakness baseline, and ultimately outputs security \weaknesses of the app.}
Figure~\ref{fig:architecture} shows the overview of \ausera, including three phases: 
(1) \textit{Sensitive data tagging}, which identifies sensitive data in \apps, including user inputs and the data from server displayed 
{on} the UI pages, and then attaches semantics to the sensitive data-related variables in .xml/.java files according to our constructed sensitive keyword database.
(2) \textit{Function identification}, 
\revised{which identifies the functions related to data leakage such as preference storage, SMS transmission, and determines the behavior of a piece of code based on API invocations (or their call sequence patterns).} 
(3) \textit{\Weakness detection}, which performs taint analysis based on the tagged sensitive data and functions to check the existence of weaknesses in the proposed baseline.

\begin{figure}
	\centering
	\includegraphics[width=0.5\textwidth]{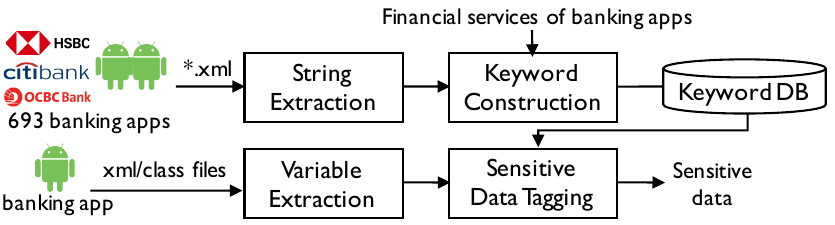}
	\caption{\sen{Identification of sensitive data}}
	\label{fig:sendata}
\end{figure}

\noindent $\bullet$ {\bf Sensitive data tagging.}\label{sec:approach:tagging}
{Since we are concerned about the sensitive data in banking apps that may incur security risks,}
we manually extract {typical} data-related keywords in banking apps.

\begin{figure}
	\centering
	\includegraphics[width=0.45\textwidth]{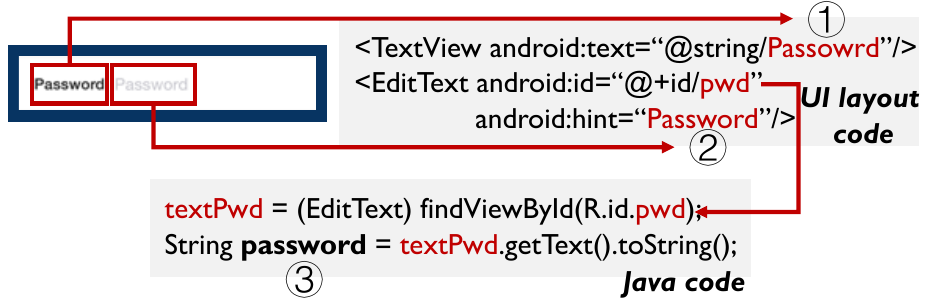}
	\caption{\sen{Code relation between Java and UI layout code}}
	\label{fig:android}
\end{figure}

\revised{Figure~\ref{fig:sendata} shows the process of sensitive keyword database construction and sensitive data tagging. }
\noindent{(1) \textit{Sensitive keyword DB construction.}}
\revised{To construct the keyword database, we first extract all strings (i.e., component ID name of {EditText}, the hint text of {EditText}, and the text of {TextView}) from the layout files of 693 \apps by reverse engineering. We then filter the strings according to the core financial services of banking apps such as login, payment, etc.}
Note that, to avoid missing variants of the keywords, we further employ \textsc{Word2vec}~\cite{mikolov2013efficient} to supplement the corpus of the keyword database. Specifically, we load the trained model based on the \emph{.bin} word vector, using the sentences extracted by \textsc{Supor}~\cite{huang2015supor} from  54,371 general apps.
For example, we further find the string ``passwd'' is a variant of the sensitive keyword of ``password.''
These sensitive keywords are able to indicate the semantics of the components (i.e., {EditText} and {TextView}).
\revised{Eventually, we build a sensitive keyword database containing 70 keywords, which can be classified into 4 categories as shown in Table~\ref{tbl:keyword_dataset}.
	Currently, we only consider two languages (i.e., English and Chinese). In the future work, we may extend the language types.
	The full list of keywords is publicly available online.\footnote{\url{www.sites.google.com/view/ausera/}}	
}
\noindent{(2) \textit{Sensitive data tagging.}}
\revised{Based on the keyword database, we can identify the sensitive data-related variables in the code and attach semantics to them.}
\revised{Specifically, we first extract variables related to two kinds of components: {EditText} for user input and {TextView} for data display. For each component, as shown in Figure~\ref{fig:android}, there may be several variables declaring different aspects of the component such as the component ID, component hint, and component text. Therefore, we extract all the variables related to each component, and then tag the variable as sensitive if it matches with any keyword in the keyword database.
	{Note that, the semantic tagging method in {previous work}~\cite{huang2015supor,andow2017uiref} relies on the component relation in layouts, which may {lose} some user inputs.}
	As a result, sensitive data is tagged with its semantics in the format $\langle$\emph{variable}, \emph{keyword}$\rangle$.}

{For} the example described in the introduction, the sensitive data is tagged as $\langle$edit\_PIN, pin$\rangle$, $\langle$edit\_firstName, firstname$\rangle$, $\langle$edit\_lastName, lastname$\rangle$, $\langle$edit\_addr, addr$\rangle$, so 
{the app in the example} is confirmed to send sensitive data out via SMS. 

\begin{table}[t]\small
	\centering
	\caption{Keyword examples}
	\label{tbl:keyword_dataset}
	\begin{tabular}{l l c}
		\hline
		{\bf Category} & {\bf Keyword Examples} & {\bf Number}\\
		\hline
		{\bf Identity}  & username, userid, byname, user-agent & 13  \\ \hline
		{\bf Credential}  & password, passcode, pwd, pin & 11\\ \hline
		{\bf Personal Info}  & name, phone, email, birthday & 24 \\ \hline
		{\bf Financial Info}   & credit card, amount, payment, payee & 22\\ \hline
	\end{tabular}
\end{table}

\noindent $\bullet$ {\bf Function identification.}\label{sec:approach:function}
%
{The sensitive data extracted above are defined as \emph{sources}, and be far apart from the access of unauthorized users. We use our newly-defined sinks (i.e., specific-APIs) to identify function code that is associated with \weaknesses for banking services.
}
However, as discussed in Section~\ref{sec:baseline}, these sensitive data may be divulged during the storage or transmission process. To achieve confidentiality, the sensitive data should not flow into a code point where unauthorized users can access via local storage, external storage, logging output, {SMS}, and component transition in Table~\ref{tbl:issues} (\aka, \textit{sinks} of sensitive data). 
\sen{It is worth mentioning that the sinks here are different from the sinks defined in \textsc{SuSi}~\cite{rasthofer2014machine}.
\textsc{SuSi}'s sinks are all potential method calls with 12 categories that leak sources out of mobile devices, while our newly-defined sinks are leaking sensitive data through specific leakage channels (e.g., shared preferences, logging output, and SMS).}
According to the leakage channels, we manually define 106 vulnerable sinks~\cite{ausera} in total that are likely to be exploited in \apps.

Communication infrastructure, which is indispensable to banking apps~\cite{reaves2015mo,egele2013empirical}. It establishes a channel to communicate with remote bank servers. However, communication infrastructure is likely to be attacked, and hence it can undermine the security of these apps. The core functionalities in communication infrastructure include certificate verification, cryptographic operation, and host authentication. To accurately identify the functional code for communication infrastructure, we summarize all invocation patterns of multiple Android APIs for each functionality. Taking hostname verification as an example, if there is an invocation sequence \{\texttt{new X509HostnameVerifier}, \texttt{setHostnameVerifier} of class \texttt{HttpURLConnection}\}, we consider that the app 
{uses} hostname verification during communication. \revised{We further check its implementation to determine whether it implements correctly.
We have 12 groups of API invocation patterns in total for function identification in communication infrastructure.} We reverse-engineer banking apps, locate the invocations of these relevant Android APIs, and use call graphs and component transitions to determine their call relation in between.
Finally, we can identify the functional code for communication infrastructure of \apps.

\noindent $\bullet$ {\textbf{Security weakness detection.}}\label{sec:approach:detection}
Given a banking app, we attempt to find whether it contains any \weaknesses listed in Table~\ref{tbl:issues} and reduce false positives by employing the two strategies:  
\sen{\emph{(1) a forward data-flow analysis to determine whether there exists sensitive data flowing into insecure sinks by leveraging sensitive data tagging and taint analysis; (2) a backward control-flow analysis to check whether the vulnerable functional code identified by API invocation patterns in communication infrastructure is feasible based on call graphs and Activity transitions.}}

We carry on a forward taint analysis on top of \textsc{Soot}~\cite{vallee1999soot} to support intra- and inter-component communication analysis based on the tagged sensitive data.
These data are regarded as \textbf{\emph{sources}}, and the sinks are the Android API list we defined.
During the process of functional code identification, we can obtain all vulnerable code \sen{(i.e., \textbf{\emph{incorrect implementation}})} that exists in communication infrastructure. However, noise may 
{arise} because the dead code for testing purpose cannot be executed during runtime. Reaves \etal~\cite{reaves2015mo,reaves2017mo} found that the dead code may bring false positives to the detection results. We perform a backward control-flow analysis, and extract all reachable call sequences according to call graphs and Activity transitions. If the vulnerable code is reachable, we determine it is a valid \weakness, \revised{otherwise, it is a false alarm}.

We highlight the following three strategies to reduce false positives.
(1) \ausera reduces the size of our extracted keywords from 124 to 70, which effectively reduces ambiguity of the keywords (\eg, ``info'' and ``status''), and hence can identify sensitive data more accurately.
(2) \ausera utilizes newly-defined sources and sinks, which are relevant to \weaknesses of sensitive data leakage.
(3) \ausera identifies the vulnerable code and checks its reachability to eliminate \revised{weaknesses} in dead code by call graphs and Activity transitions.

\subsubsection{\textbf{Implementation of \ausera}.}
To implement \ausera, we combine static program analysis and sensitive data tagging to identify sensitive data in banking apps, and associate them with the corresponding variables in XML/Java code. 
\ausera relies on \textsc{Apktool}~\cite{apktool} to extract resource files from apks. It then uses parts-of-speech (POS) tagger of~\textsc{OpenNLP}-1.8.3~\cite{nlp} to parse the text labels in {TextView} and {EditText}, thereby identifying keywords included. We manually check on these keywords to retain the ones that are sensitive and relevant to the core functionalities of banking apps. After that, we employ \textsc{Word2vec} to supplement the keyword database. 

To accomplish the detection, we summarize 12 groups of  patterns (\eg, {AES/ECB/NoPanding}) to depict the communication \weaknesses. Then we employ 
{pattern-based} static analysis to find the possible vulnerable patterns in code. 
We check three aspects for certificate authentication: whether the client side 1) allows all hostname requests; 2) bypasses hostname verification; 3) fails to implement anything in the server verification method (\texttt{checkServerTrusted}).
The \weakness ``hard-coded encryption key'' is determined by first checking whether an encryption key is embedded in code, and examining whether it is used to encrypt sensitive data to reduce false positives. 
The banking sensitive data are encrypted with the {DES} or {Blowfish} algorithm. Using either of the encryption mechanisms is viewed as a \weakness~\cite{reaves2015mo, reaves2017mo}. The {AES} forbids ECB mode because it does not provide a general notion of privacy~\cite{egele2013empirical}. 
The padding of {AES} and {RSA} is always improper, such as \texttt{NoPadding} and \texttt{PKCS1}, though {AES/ECB/NoPadding} is very frequently used. The function SecureRandom should not be seeded with a constant. The hash functions {MD5} and {SHA-1} are insecure~\cite{wang2005break,wang2005finding}.

\subsubsection{\textbf{Evaluation of \ausera}}\label{sec:eval:baseline}
We randomly selected 60 \apps (12.8\%) in our dataset and manually checked the detection results to evaluate \ausera's precision.
\revised{Note that, we cannot evaluate the false negatives when assessing banking apps due to lack of weakness benchmarks of banking apps.}
\revised{False positive (FP) refers to weakness that are detected during static analysis but actually infeasible at runtime or detected by mistake.} As a result, we only found 6 false positives (corresponding to five weakness types, \ie, Shared Preference Leakage, Logging Leakage, SD Card Leakage, Text File Leakage, and Hard-coded Key) from the identified 341 \weaknesses of these 60 \apps, \revised{achieving} an average precision of 98.24\%.

Consequently, 5 out of 6 false positives belong to sensitive data leakage. The reason is that \ausera matches variables (\eg, ``{pkgname.txt},'' ``{login\_fragement},'' ``{loginpager},'' and ``{spinnerGender}'') inaccurately with the keywords in our database. \revised{The remaining one FP belongs to Hard-coded Key type, because the extracted variable is relevant to the exception parameters (\ie, ``{KeyPermanentlyInvalidateException}'').}

\begin{table}[t]\small
	\centering
	\caption{Distribution of the collected banking apps}
	\label{fig:subject_dataset}
	\begin{tabular}{lcccc}
		\hline
		\textbf{Continent} & \textbf{\#Developed } & \textbf{\#Developing} & \textbf{Total} & \textbf{{Percentage}} \\ \hline
		{\bf Europe}        & 102         & 0                & 102      &  21.7\%  \\ 
		{\bf America}       & 53          &24                & 77       &  16.4\%  \\  
		{\bf Asia}          & 16          & 210              & 226      &  48.1\%  \\ 
		{\bf Oceania}       & 16          & 0                & 16       &  3.4\%  \\ 
		{\bf Africa}        & 0           & 49               & 49       &  10.4\%  \\ 
		\hline
		{\textbf{Total}} & {187}      & {283}            & {470}     & - \\ \hline
	\end{tabular}
\end{table}

\subsection{RQ1: Tool Evaluation and Data Collection}\label{sec:evaluation}

\noindent \textbf{Banking app collection.}
As shown in Figure~\ref{fig:overview},
we collected 693 banking apps\footnote{{Apart from the 693 apps, we have filtered out apps with packer techniques (49 apps in total) and with decompilation failure since they are out of scope in this paper.}} in total from various Android markets such as Google Play store and APKMonk~\cite{apkmonk}. 
%
{Note that we only collect multiple versions of some apps from APKMonk to conduct the longitudinal anlaysis (cf. Section 3.3) since APKMonk maintains the full versions of apps, while Google Play store only maintain the latest version.}
The collected apps range across 470 unique banks, where some apps have multiple versions.
They originate from both developed and developing countries across five continents (see breakdowns in Table~\ref{fig:subject_dataset}). 
Table~\ref{fig:subject_dataset} indicates that 48.1\% of the banking apps are from Asia, considering the largest population proportion all over the world. Only 3.4\% of apps are from Oceania,  considering its smallest population proportion all over the world.
The 24 \apps of American developing countries all originate from South America, while 16 apps of  Oceanian developed countries originate from Australia and New Zealand. 16 apps of Asian developed countries originate from Singapore, Japan, and South Korea. To our knowledge, this is the largest \app dataset taken into study to date.

\noindent\textbf{Comparison with the state of the practice}.
We compare the detection results of \ausera with 4 industrial and open-source tools, including \textsc{Qihoo360}, \textsc{AndroBugs}, \textsc{MobSF}, and \textsc{QARK}. We randomly select 60 \apps in our dataset for comparison and run each tool 3 times to stabilize the  detection accuracy. \revised{Table~\ref{tbl:comp} shows the results.} 
\ausera outperforms other tools in both precision and time cost, achieving 98.24\% precision in 1.6 minutes per app.
The precisions for these tools are obtained by manual validation through filtering out all false positives. 
\revised{We also conduct a cross-validation of detection results across different authors.}
We can see that all comparisons of the detection results comply with the \weaknesses baseline. 
\revised{\ausera outperforms other tools with higher precision and less time. {\ausera} manages to scan each app within 1.6~minutes on average, much faster than the other tools.} 

\begin{table} 
	\small
	\caption{Detection result comparisons}
	\vspace{-2mm}
	\label{tbl:comp}
	\begin{minipage}{\columnwidth}
		\begin{center}
			\begin{tabular}{c c c c}
				\hline
				{\bf Tools} & {\bf \#Types} &{\bf Precision} & {\bf Time/App (mins)}\\
				\hline
				\ausera                      & 341 & 98.24\% & 1.6 \\ 
				\hline  \textsc{Qihoo360}  & 80 & 87.50\% & 8.5 \\ 
				\hline  \textsc{AndroBugs} & 76 & 81.58\% & 1.8 \\ 
				\hline  \textsc{QARK}      & 93 & 87.10\% & 16.1 \\ 
				\hline  \textsc{MobSF}     & 213 & 48.36\% & 2.4 \\ 
				\hline
			\end{tabular}
		\end{center}
	\end{minipage}
\end{table}

In particular, we show several specific cases to explain \revised{how false positives are incurred}.
Sensitive data disclosure through logging \revised{can be} detected by \textsc{MobSF}, 
however, \textsc{MobSF} just matches the following APIs if used (\eg,  \texttt{Log.e()}, \texttt{Log.d()}, and \texttt{Log.v()}), \revised{without further determining whether the output data is sensitive or not}. There is no doubt that it \revised{would incur a large number of} false positives. If the data is not sensitive, such as ``\texttt{menu\_title},''  \revised{it is very normal for developers to output it in the terminal or write messages to understand the state of their application.} The risk is that some credentials (\eg, PIN and password) are also leaked by logging outputs.
A syntax-based scanning tool \revised{may} provide an incomplete and incorrect analysis result due to the influence of dead code. For example, \revised{\textsc{Qihoo360}} detected three code blocks violating server verification, \eg, do nothing in \texttt{checkServerTrusted}. In contrast, \ausera aims to 
\revised{to minimize the influence of dead code.}
Two key strategies to eliminate such false positives are: (\rmnum{1}) checking whether invalid authentication is in a feasible path in call graphs; (\rmnum{2}) checking whether the \texttt{Class} has been instantiated in Activity transitions.	

\revised{Apart from the comparison with the above 4 tools, we also discuss the comparison between \ausera and two taint analysis tools (i.e., \textsc{FlowDroid}~\cite{arzt2014flowdroid} and \textsc{IccTA}~\cite{li2015iccta}). 
\ausera aims to identify weaknesses specifically in \apps, while \textsc{FlowDroid} and \textsc{IccTA}, which largely rely on sources and sinks defined in \textsc{SuSi}, aim to identify the data leakage in general apps. 	
	(1) The sources and sinks considered by \textsc{FlowDroid} and \textsc{IccTA} are specified by \textsc{SuSi}, which contains 12 different source categories and 15 different sink categories. However, among them, we only use taint analysis on 4 types of weaknesses (i.e., Shared preference leakage, logging leakage, SD card leakage, and SMS leakage). 
	In other words, \textsc{FlowDroid} and \textsc{IccTA} cannot detect most of security weakness types in our proposed data-related baseline specific to \apps. (2) 
	{In fact, we have deployed \textsc{FlowDroid} and \textsc{IccTA} on our defined sources and sinks, and find that they} cannot identify the concrete data types (i.e., sensitive or non-sensitive) when tracking the 4 types \revised{of weaknesses}. For example, developers usually output debug information such as string length via logging channel, however, tracking such non-sensitive data causes many false positives. While \ausera only tracks the labeled sensitive data that are most relevant to the core financial services of banking apps.
	(3) Most of the sources defined in \textsc{SuSi} are not sensitive in banking apps, such as the API invocations of Bluetooth, Calendar, and Settings.}
\revised{More comparison results can be found 
{on} our website~\cite{ausera}.}

\vspace{1mm}
\noindent\fbox{
	\parbox{0.95\linewidth}{
		\textbf{Answer to RQ1.} In summary, existing state-of-the-practice tools are less 
		{effective} (i.e., lower precision, more false positives, and cost more time) in identifying data-related weaknesses in banking apps, compared with \ausera.
		Therefore, \ausera can be used to collect a large number of security weaknesses for further in-depth analysis.
	}
}

\vspace{1mm}
\noindent{\bf Weakness collection.}
\ausera is demonstrated as the most effective tool to collect banking specific security weaknesses, we thus apply it on the collected 693 banking apps across 83 countries. Finally, we collect 2,157 security weaknesses for further large-scale empirical study.

\begin{table}\small
	\renewcommand{\arraystretch}{1.1}
	\centering
	\caption{\Weaknesses in 470 banking apps}
	\vspace{-2mm}
	\label{tbl:number}
	\resizebox{0.45\textwidth}{!}{
		\begin{tabular}{llc}
			\hline
			{\begin{tabular}[c]{@{}l@{}} \textbf{\Weakness Category}\end{tabular}} & \textbf{\Weakness Type} & \textbf{\#Affected Apps} \\
			\hline
			\begin{tabular}[c]{@{}l@{}} \textbf{Input Harvest}\end{tabular} & Screenshot & 415 (88.3\%) \\ \hline
			\multirow{5}{*}{\begin{tabular}[c]{@{}l@{}}\textbf{Data Storage}\end{tabular}}
			& {Shared preference} & 44 \\
			& WebView DB & 64 \\
			& Logging & 66 \\
			& SD Card & 14  \\
			& Text File & 10 \\ \hline
			\multirow{2}{*}{\begin{tabular}[c]{@{}c@{}}\textbf{Data} \\ \textbf{Transmission}\end{tabular}} & {SMS} Leakage & 18  \\
			& ICC Leakage & 324 (68.9\%)  \\ \hline
			\multirow{8}{*}{\begin{tabular}[c]{@{}l@{}} \textbf{Communication} \\ {\bf Infrastructure} \end{tabular}} & HTTP Protocol & 84  \\
			& Invalid Certificate & 31  \\
			& Invalid Authentication & 222  \\
			& Hard-coded Key & 30  \\
			& Improper AES & 131  \\
			& Improper RSA & 231  \\
			& Insecure SecureRandom & 133  \\
			& Insecure Hash Function& 340 (72.3\%) \\ \hline
	\end{tabular}}
\end{table}

\section{A Large-scale Comprehensive Empirical Study}\label{sec:study}
In this section, we conduct a large-scale empirical study from different aspects based on the collected weaknesses by \ausera.

\subsection{RQ2: Security Status of Banking Apps}
\revised{Since the multiple versions of a banking app may have overlapped weaknesses, we select the latest version of the 693 apps if they have multiple versions, and apply \ausera to} these 470 unique banking apps to 
conduct the following study.
Table~\ref{tbl:number} shows the results of \weaknesses corresponding to the security baseline defined in Section~\ref{sec:baseline}.

\noindent {\bf Input harvest.}
Screenshot (88.3\%), as an easy-to-use way to harvest users' credentials, is most likely to be neglected by developers. Only 55 apps (\eg, Bank of Communications of China) are protected from screenshots in our investigation.

\begin{figure}
	\centering
	\includegraphics[width=0.4\textwidth]{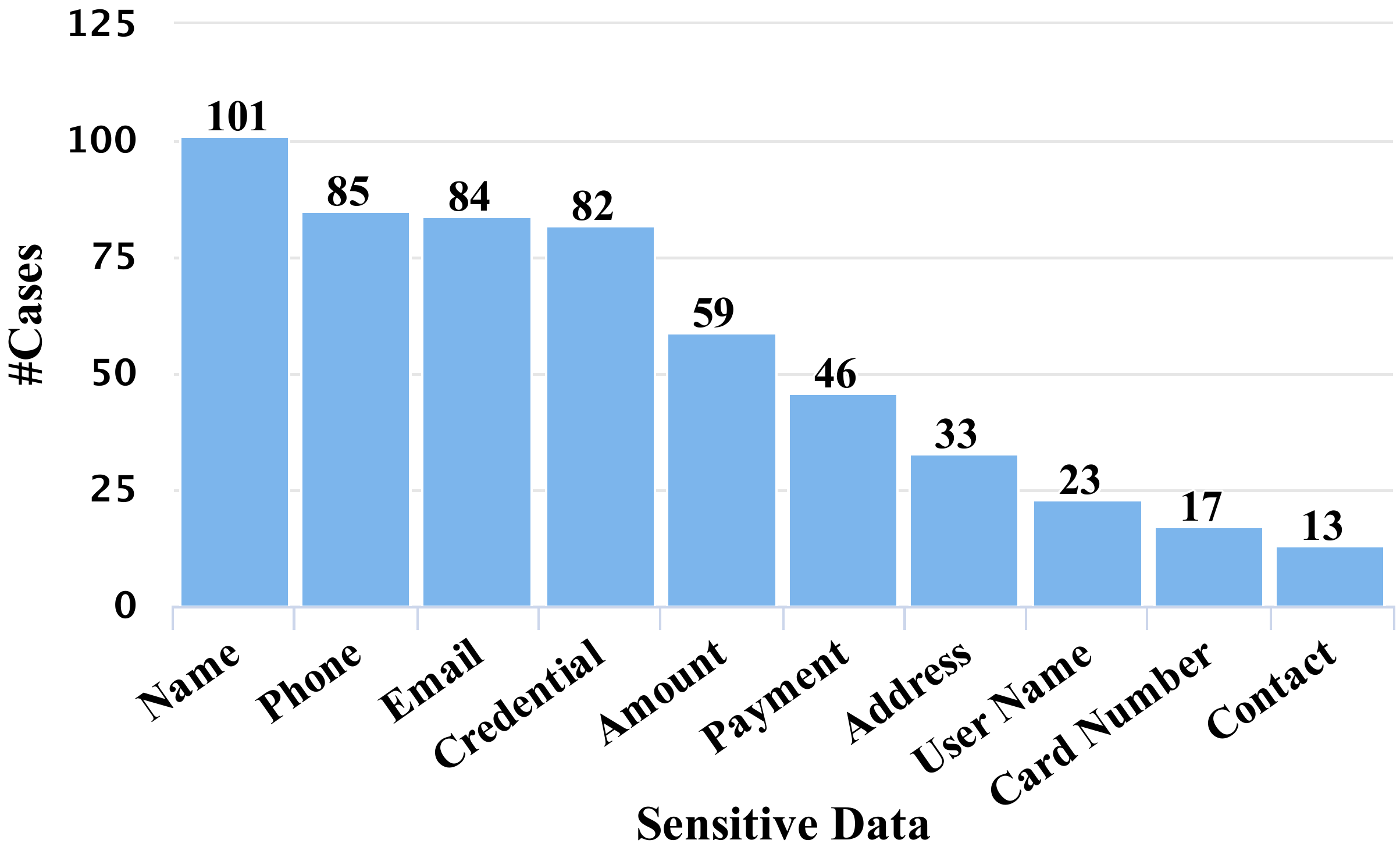}
	\caption{\sen{\small Top 10 sensitive data types leaked in 470 banking apps}}
	\label{fig:sensitive_data}
\end{figure}

\noindent {\bf Data storage.} Only a small portion of apps store sensitive data on SD Card (2.98\%) and Text File (2.13\%), which are globally accessible and thereby susceptible to privacy leakage. 
\sen{We show that {Shared preference}, {Logging}, and {WebView DB} are the main channels that leak sensitive data.}
\sen{As shown in Figure~\ref{fig:sensitive_data}, \ausera identifies 592 cases of private data leakage across 470 unique \apps. \emph{Credentials} (\eg, PIN), as the most dangerous leakage in banking apps, appear in 82 cases and affect 64 banking apps. Note that \emph{banking-specific data} (\eg, transaction password and card number) accounts for 22.47\%, and the other data leakage includes \emph{personal info} (\eg, Name, Phone, and Email).}

\noindent {\bf Data transmission.} We show that ICC Leakage (68.9\%) is also among the most popular \weaknesses. Despite the {small} portion of {SMS} Leakage, {SMS} could directly forward credentials, thwarting confidentiality.  
\revised{For example, the real banking app mentioned in the introduction leaks sensitive data such as pin, first name, last name, and address via SMS.}

\noindent {\bf Communication infrastructure.} The protection of communication infrastructure in \apps is far away from satisfactory. More specifically, many apps are still using {HTTP} to exchange sensitive data with the remote bank server, or do not validate the certificates of the connected servers.
We find 222 banking apps with invalid authentication, including 13 \apps that have both invalid and correct {SSL/TLS} implementations in source code. They establish communications with servers using different strategies \sen{(i.e., invalid and correct SSL/TLS implementation)}.
Insecure Hash Function (72.3\%) is also frequently misused. 

\vspace{1mm}
\noindent\fbox{
	\parbox{0.95\linewidth}{
		\textbf{Answer to RQ2.} Overall, the security status of banking apps is severe according to the results. In summary, Screenshot (88.3\%), Insecure Hash Function (72.3\%), and ICC Leakage (68.9\%) are the most popular \weaknesses of \apps. Meanwhile, Invalid Authentication (222 apps) also has severe damage.
	}
}

\subsection{RQ3: Global Distribution of Weaknesses}\label{sec:finding:demography}
\revised{Figure~\ref{fig:map} shows the number of \weaknesses discovered among the banking apps by continents.}
The 
\revised{intensity scale}
encodes the number of \weaknesses the apps have, scaled from light blue (least) to dark blue (most). 
\revised{We observe the following findings:}
(1) \Weaknesses \revised{in} \apps of Asia outnumber those of Europe (resp. North America) by 1.56 (resp. 1.31) to 1, where each \app of Asia has 6.4 \weaknesses on average, indicating that the \apps of developed countries (\ie, Europe and  North America) have fewer \weaknesses than those of developing countries. Ironically, to our surprise, we find that \weaknesses \revised{in} apps of Asian developed countries slightly outnumber (with 6.7 \weaknesses per app) those of Asian developing countries. 
(2) \Apps from Africa exhibit satisfactory security status, having only 4.6 \weaknesses on average, some are even more secure than those of developed countries.
\revised{Possible} reasons \revised{why} the security of banking apps varies across regions can be interpreted as follows: 

\begin{figure}
	\centering
	\includegraphics[width=0.4\textwidth]{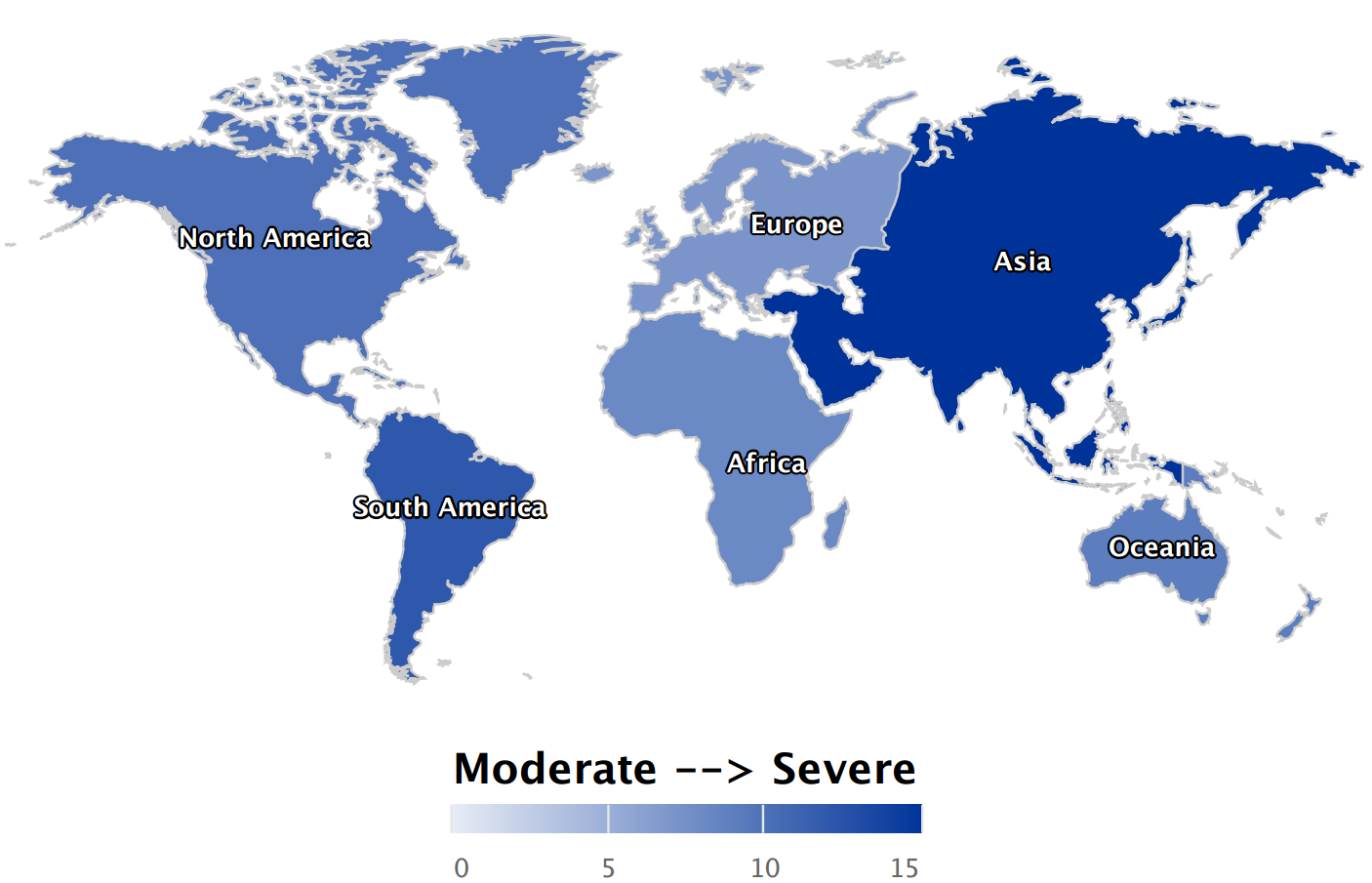}
	\caption{Number of weaknesses in global banking apps}
	\label{fig:map}
\end{figure}

\begin{itemize}[noitemsep,topsep=0pt,leftmargin=*]
	
	\item The financial regulations and development guidelines are different across regions, which may affect the implementation \revised{of banking apps}. For example, both Europe (GDPR~\cite{gdpr}) and USA (PCI DSS~\cite{pci}) adopt very strict security and privacy regulations. The GDPR poses a regulatory framework that is unique to the financial service industry. Failure to meet its requirements will come with potentially hefty penalties~\cite{EU:95}. This is also reflected by the 143 banking apps from Europe and USA, where data leakage rarely exists, with only 0.27 data leakage \weakness reported per app.
	
	\item The development budget and developers' expertise may affect the security of products. During our investigation, we find that a number of local banking apps of China have many more \weaknesses than international  or nationwide ones. 
	We speculate that due to inadequate budget for app development, those released apps are prone to being less secure.
	
	\item Cashless payment systems have been bootstrapped in areas where traditional banking is uneconomical and expensive, 
	\revised{removing}
	large \revised{investments} on the massively deployed financial infrastructure.
	This is evidenced by the fact that Kenya, a country in Africa, 
	\revised{is a world leader of}
	money transfers by mobile~\cite{firstmoney}, and 68\% \revised{of} people in Kenya report \revised{the} use of phones for a financial service~\cite{Kenya}. 
	
\end{itemize}

\vspace{1mm}
\noindent\fbox{
	\parbox{0.98\linewidth}{
		\textbf{Answer to RQ3.} We conclude that apps across different countries exhibit various types of security status, mainly because of different economies and regulations that take shape.  We find that apps \revised{from} Africa have comparatively moderate security status, primarily because of its high demand for cashless services. 
	}
}

\subsection{RQ4: Longitudinal Analysis of Version Updates and Fragmentation}\label{sub:longitudinal}
We 
{attempt} to perform a longitudinal study on security risks by revisiting the 7 apps (GCash, mPay, MOM, Zuum, Oxigen Wallet, Airtel Money, and mCoin) which have been systematically studied by Reaves \etal~\cite{reaves2015mo}, with confirmed \weaknesses.
We downloaded all available versions of 6 apps (mCoin is excluded since history versions are not publicly available.), \revised{and obtained 88 different versions in total}, \ie, GCash (6 versions), mPay (20 versions), MOM (22 versions), Zuum (12 versions), Oxigen Wallet (12 versions), and Airtel Money (8 versions). All versions span more than two years.

\begin{figure}
	\centering
	\includegraphics[width=0.47\textwidth]{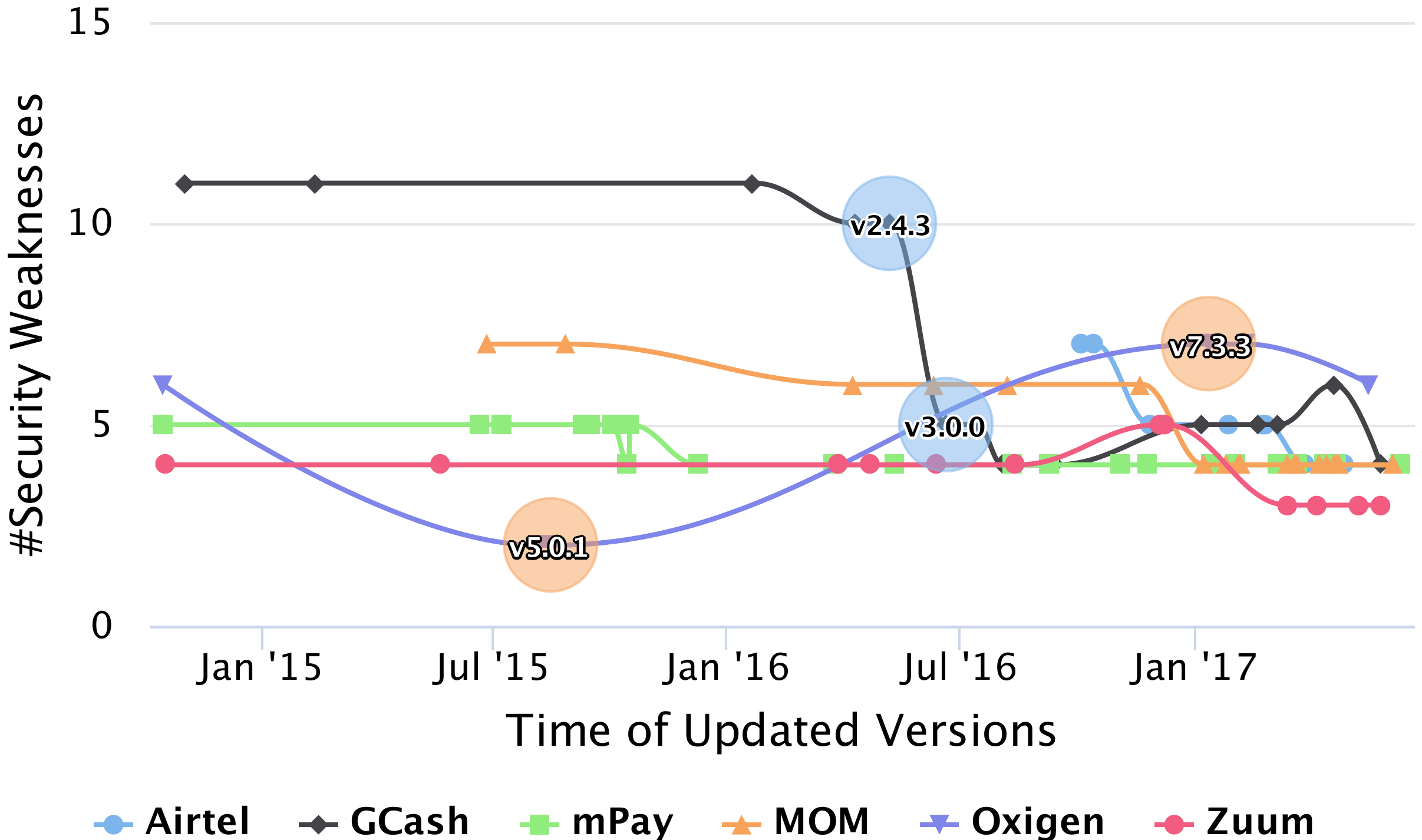}
	\caption{Number of \weaknesses in each update version.
	}
	\label{fig:version}
\end{figure}

Figure~\ref{fig:version} shows the number of detected \weaknesses across all versions of each app.
We can see most of the version updates (90\%) fail to bring at least two successful patches for \weaknesses in their history versions, which echoes the \revised{findings of} paper~\cite{reaves2017mo} that apps have not repaired critical vulnerabilities in their new versions. After an in-depth manual analysis, we find input harvest via screenshots, MITM attacks, {AES/RSA} misuses, and insecure hash functions are the most \revised{common} \weaknesses that remain unfixed.
Furthermore, developers usually neglect hostname verification or server authentication, which may \revised{enable} the MITM attack. These apps are also not aware of {AES/RSA} misuses and insecure hash functions, indicating that developers are still not aware of these \weaknesses perpetually.

GCash has a sharp decline from v2.4.3 to v3.0.0 in terms of the number of \weaknesses.
Three \weaknesses are patched, the hard-coded encryption key, insecure SecureRandom, and privacy leakage to SD Card. Reaves \etal~\cite{reaves2015mo,reaves2017mo} found that \revised{the vulnerabilities still remain in the updated version in 2016. However, according to our security reports, GCash fixed most of the vulnerabilities in their latest version.}
In contrast, the \weaknesses of Oxigen Wallet significantly increase from v5.01 to v7.3.3 due to the changes of app features.
More specifically, many new \weaknesses (\ie, \texttt{WebView DB} Leakage, ICC Leakage, MITM Attacks, and Insecure SecureRandom) \revised{were} introduced, which \revised{had} not been discovered by Reaves \etal~\cite{reaves2015mo}. They compared the code similarity between the 2015 and 2016 versions of each app, and found some apps have significant \revised{amounts of} new code~\cite{reaves2017mo}.
This 
\revised{aligns} with our study that many banking apps 
\revised{do not perform} systematic security checks before delivery. 

\begin{figure}
	\centering
	\includegraphics[width=0.4\textwidth]{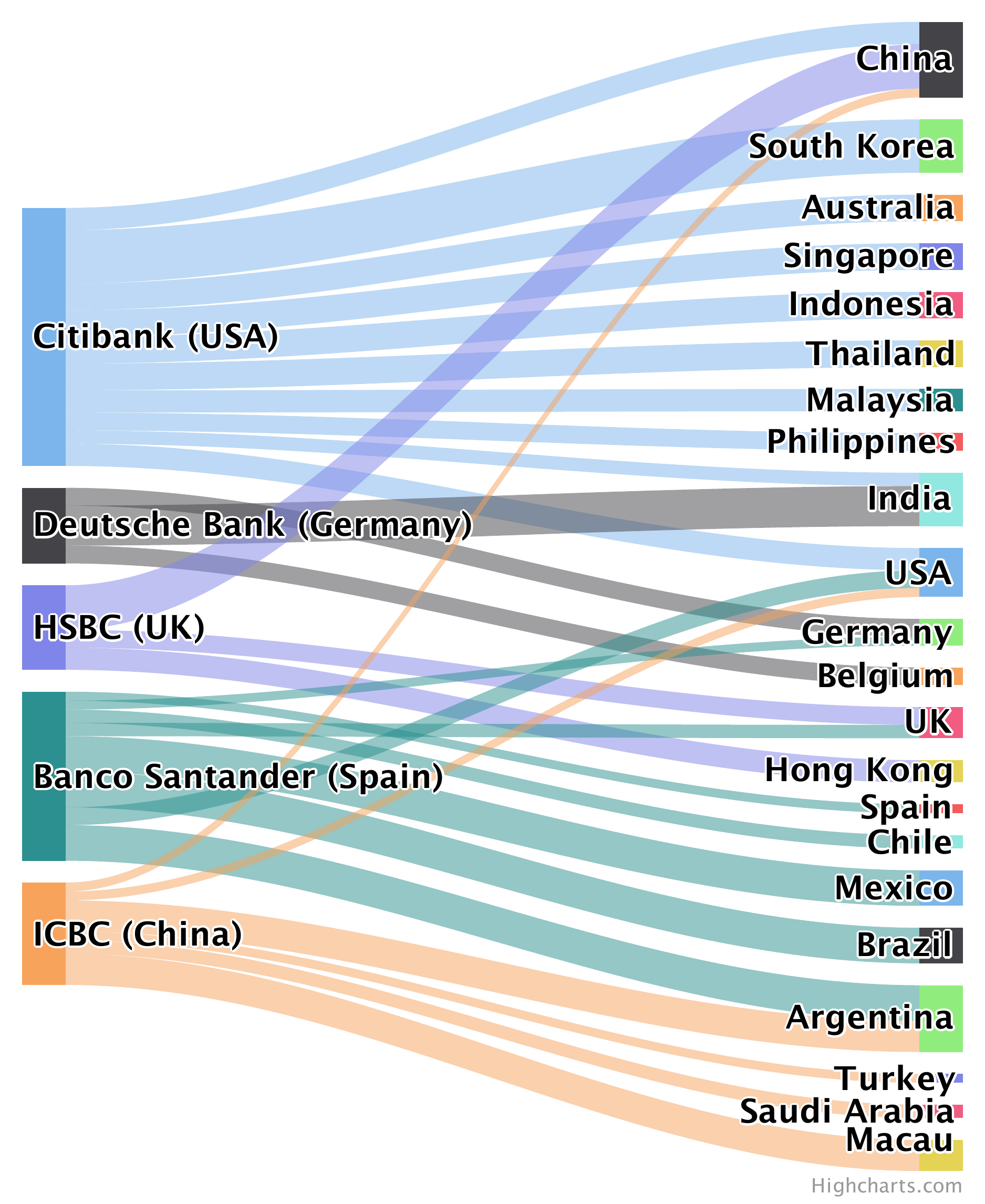}
	\caption{Flow of security \weaknesses from parent banks to subsidiary banks across the world. The flow width indicates the number of \weaknesses originating and terminating between two corresponding banks. Different types of banks are encoded by different colors.}
	\label{fig:mul_versions}
\end{figure}

Furthermore, we find banks encounter the version fragmentation problem especially when they release versions to different markets by countries. We selected the top 5 \apps based on the S\&P Global Market Intelligence report~\cite{top100} across their 30 different versions, \ie, Citibank (10 versions), HSBC (3 versions), Deutsche Bank (3 versions), Banco Santander  (8 versions), and ICBC (6 versions). By comparing the differences of \weaknesses between these versions, we observe the following:
(1) A subsidiary bank, incorporated in the host country but owned by a foreign parent bank, usually  launches its original financial services with most of its products, such as \apps,  into the host market. 
As a result, a subsidiary bank inherits the \weaknesses  from the original version of its parent bank. This observation is evidenced by the South Korean version of Citibank app and the Macau version of ICBC app (see Figure~\ref{fig:mul_versions}).
(2) Due to the business difference, culture difference, and expertise of security teams,  \weaknesses of apps vary across different markets by countries. This is also evidenced by the fact that the official app of HSBC (China) v2.7.1 has more \weaknesses 
than that of HSBC (UK) and HSBC (Hong Kong). 
\revised{A possible reason might be that} HSBC (China) is independent of the parent bank in terms of its app development outsourcing procedures and security teams, while in Hong Kong, as the former UK colony, HSBC (Hong Kong) largely follows the convention of HSBC (UK). Nevertheless, we find that not all subsidiary banks operate under the host country's regulations in terms of the number of \app security risks (Figure~\ref{fig:mul_versions} shows the source and host countries of flows containing security \weaknesses.).

\vspace{1mm}
\noindent\fbox{
	\parbox{0.95\linewidth}{
		\textbf{Answer to RQ4.} By revisiting apps studied by previous research and further examining them across all their publicly available versions that have not been scrutinized before, we conclude that app developers are still not aware of these \weaknesses perpetually. Furthermore, apps owned by subsidiary banks are always less secure than or equivalent to those owned by parent banks, for which the assumption that subsidiary banks operate under the host country's regulations does not always hold true.
	}
}

\subsection{RQ5: Weakness Fixing and Feedback}\label{sub:feedback}
Our study has uncovered \all weaknesses \revised{in total} from 693 \apps, most of which have been reported to the corresponding banks.  
As shown in Table~\ref{tbl:reply}, \feedback banks have replied and confirmed these weaknesses, and \textbf{\patched apps} have been patched.\footnote{\revised{
We do not 
\revised{disclose} any concrete weakness types or details in these banking apps to avoid security threats.}}
Furthermore, we approached the major stakeholders across the global, such as HSBC (UK/Hong Kong/Shanghai), OCBC (Singapore), DBS (Singapore), and BHIM (India), to understand their security practice and policies. Through in-depth discussions with 7 banks, we find they hold different mindsets toward assessing severity of \weaknesses and setting security goals. 
{Note that, on average, the 7 banks take 41 days to fix their security weaknesses we reported.}
We elaborate this gap and provide our insights on how to close it.

{\noindent {\bf {Lack of effective criteria for rating security weaknesses}}}.
An effective severity criterion of \weaknesses is crucial for banks to prioritize security patching.
However, such a criterion is still missing for banking apps. As a result, some banks use \textsc{CVSS}~\cite{cvss} to determine the severity of the identified \weaknesses. However, this standard is not perfect in practice~\cite{munaiah2016vulnerability,allodi2018identifying,notcvss_3,notcvss_4}, and provides few principled ways to characterize security risks and potential impact. 
Moreover, we find these banks hold subjective attitudes toward fixing different types of weaknesses. For example, most banks 
{are concerned about} obvious privacy leakage (\eg, leakage from \texttt{SharedPreference}, Logging, {SMS}, SD Card,  Text File, and WebView DB), while they are only aware of and somehow reluctant to fix the weaknesses, such as ICC Leakage, Invalid Certificate, and Insecure Hash Function. 
Table~\ref{tbl:confirm} summarizes our observations on various banks' attitudes towards different weaknesses, which are classified by ``Concerned" (high priority) and ``Aware" (low priority).

{\noindent{\bf {Lack of systematic security checks and validation tools.}}} Many banking apps do not undergo a systematic security check and validation before delivery --- 
\ausera discovers a large number of high-severity \weaknesses, \eg, sensitive data leakage, hard-coded key and invalid authentication. With the assistance of \ausera, many banks, \eg, OCBC and Zijin Bank, expeditiously patched the weaknesses in their new versions.
However, ironically, some banks patched the weaknesses but introduced new ones at the same time.
For example, 
C$\ast$
patched two weaknesses (\ie, Logging Leakage and {HTTP} Protocol) by employing {SSL} over {HTTPS} communication. However, new \weaknesses are introduced in the 
{updated} version, \ie, the app fails to verify the identity of the bank server (\texttt{checkServerTrusted}), which echoes the finding of \cite{reaves2015mo,reaves2017mo} that 4 apps have new vulnerabilities.
Due to lack of systematic security checks and validation tools, many security weaknesses still reside in these apps. 

\begin{table}[t]\small
	\centering
	\caption{Different concerns from banks}
	\label{tbl:confirm}
	\begin{tabular}{l p{5.9cm}}
		\hline
		& {\bf Security \Weaknesses} \\ \hline
		\multirow{4}{*}{\begin{tabular}[c]{@{}l@{}}\textbf{Concerned}\end{tabular}}&  Screenshot, \texttt{SharedPreference} Leakage, Logging Leakage, {SMS} Leakage, SD Card Leakage, Text File Leakage, WebView DB Leakage, Invalid Authentication, Hard-coded Key, Insecure SecureRandom \\ \hline
		\multirow{2}{*}{\begin{tabular}[c]{@{}l@{}}\textbf{Aware}\end{tabular}}&  ICC Leakage, {HTTP} Protocol, Invalid Certificate, Improper {AES/RSA}, Insecure Hash Function\\ \hline
	\end{tabular}
\end{table}

{\noindent{\bf {Outdated versions remain in effect in the wild.}}} 
Banks usually hold the assumption that customers always 
keep their apps updated, and thus concentrate more on the \weaknesses of latest versions than those of outdated versions.
However, this assumption is \emph{not true}, considering the device fragmentation problem --- Android apps have to be compatible with more than 10 major versions of Android OS running on over 24,000 distinct device models; and it is also \emph{dangerous}, considering attackers can leverage the \weaknesses of outdated versions to mount specific attacks.
We find that most banking apps across multiple versions still remain in effect in the wild (\eg, Apkmonk~\cite{apkmonk}).
On average, these apps have 7.7 different versions, and the most fragmented app has 25 versions.
Thus, we strongly recommend banks push compulsory app updates to the customers \revised{or block access to outdated apps}, especially when high-severity \weaknesses were patched.

{\noindent{\bf {Risks from third-party libraries.}}} Our study finds the third-party libraries, \eg, \texttt{com.google.android.gms.$\ast$} and \texttt{com.facebook.$\ast$}, are widely used in \apps.
\ausera detects {BHIM} (v2.3.6) and {MyAadhar} (v1.9.3) use insecure third-party hash functions, such as {MD5} and {SHA-1}, to produce message digests, which have already been accepted as insecure~\cite{wang2005break,stevens2017sha1}. 
\revised{Banks still use these insecure functions despite being aware of the insecure, as they assume that ordinary attackers are not capable of breaking them.}
However, it is still possible for experienced attackers to mount a \revised{large}-scale attacks by exploiting these weaknesses.
\revised{Banks are liable} if they use security-weakened or poisoned third-party libraries without careful inspection. To avoid \revised{an} ``amplification effect'' caused by the \weaknesses in third-party libraries~\cite{derr2017ccs}, we strongly recommend banks to carefully inspect third-party libraries in use.

\begin{table}\renewcommand{\arraystretch}{1}
	\centering \scriptsize
	\caption{\small \Weaknesses tracking of \feedback banking apps. \patched banks have already patched their \apps, and the rest have confirmed the \weaknesses in their replies and will fix them soon in new versions.}
	\vspace{-2mm}
	\label{tbl:reply}
	\begin{tabular}{clcccll}
		\toprule
		{\bf No.} & \begin{tabular}[c]{@{}l@{}} {\bf Banking Apps}\end{tabular} & \begin{tabular}[c]{@{}c@{}} {\bf \rotatebox{0}{\#W}}\end{tabular} & \begin{tabular}[c]{@{}c@{}} {\bf \rotatebox{0}{\#Patched}}\end{tabular} & \begin{tabular}[c]{@{}c@{}}{\bf \rotatebox{0}{\#New}}\end{tabular} & {\bf Country} & {\bf Downloads} \\
		\toprule
		1 &HSBC$^*$ & 5  & 2 & 0 & UK & 5M - 10M  \\ \midrule
		2 & PSD Bank & 3  & 2 & 0 & Germany & 50K - 100K \\ \midrule
		3 & BBBank & 3  & 2 & 0 & Germany & 50K - 100K \\ \midrule
		4 & \begin{tabular}[c]{@{}l@{}}Intesa Sanpaolo \\ Mobile\end{tabular} & 5  & 2 & 0 & Italy & 1M - 5M \\ \midrule
		5 & AIB Mobile & 8  & 1 & 0 & Ireland & 5M - 10M \\ \midrule
		6 & Alma Bank & 6  & 3 & 0 & Russia & 5K - 10K \\ \midrule
		7 & Discover Mobile & 8  & 4 & 0 & USA & 10M - 50M \\ \midrule 
		8 & \begin{tabular}[c]{@{}l@{}}Citizens Bank \\  of Lafayette\end{tabular} & 2  & 1 & 2 & USA & 5K - 10K \\ \midrule
		9 & CDB & 6  & 2 & 2 & China & 5K - 10K \\ \midrule
		10 & Zijin Bank & 8   & 7 & 0 & China & 10K - 50K \\ \midrule
		11 & DBS & 10  & 0 & 0 & Singapore & 5M - 10M \\ \midrule
		12 & OCBC & 9  & 8 & 0 & Singapore & 5M - 10M \\ \midrule
		13 & MyAadhar & 4  & 0 & 0 & India & 50M - 100M \\ \midrule
		14 & BHIM & 3 & 2  & 0 & India & 10M - 50M \\ \midrule
		15 & ICICI Netbanking & 7  & 0  & 0 & India & 100M - 500M \\ \midrule
		16 & ICICI Pockets & 7  & 0 & 0 & India & 50M - 100M \\ \midrule
		17 & GCash & 11 & 8 & 0 & Philippines & 10M - 50M \\ \midrule
		18 & Bank Australia & 7 & 2 & 0 & Australia & 10K - 50K \\ \midrule
		19 & CaixaBank & 5 & 2 & 0 & Brazil & 1M - 5M \\ \midrule
		20 & BMCE Bank & 5 & 2 & 0 & Morocco & 100K - 500K \\ \midrule
		21 & NMB Mobile Bank & 4  & 0 & 1 & Zimbabwe & 10K - 50K \\ \bottomrule
	\end{tabular}
	\vspace{1mm}
	\\ ``\#W'': The number of detected \weaknesses. ``\#Patched'': the number of patched \weaknesses in update versions. ``\#New'': The number of newly-introduced \weaknesses in update versions. \\
	\revised{``Country'' means the country of bank headquarters.\\
	 ``$*$'': The HSBC Cybersecurity team has reviewed and responded that the remaining three reported ``weaknesses'' in HSBC China version (v2.7.1) are not vulnerabilities, but are features purposely retained to support market specific customer requirements.}
\end{table}

\vspace{1mm}
\noindent\fbox{
	\parbox{0.97\linewidth}{
		\textbf{Answer to RQ5.} 
		Incomplete security \revised{criterion} provides banks wide leeway to use one-sided judgment about specific security practices. We also observe that outdated versions and weaknesses from third-party libraries are all likely to be exploited. They \revised{remain} unfixed for weeks to months post-disclosure. This gap provides opportunities  for attackers to strike. 
		Understanding the gap between industrial practitioners and academic researchers help illuminate the nature of patching process. 
	}
}

\begin{figure*}
	\centering
	\begin{minipage}[t]{.29\linewidth}
		\includegraphics[width=1\linewidth]{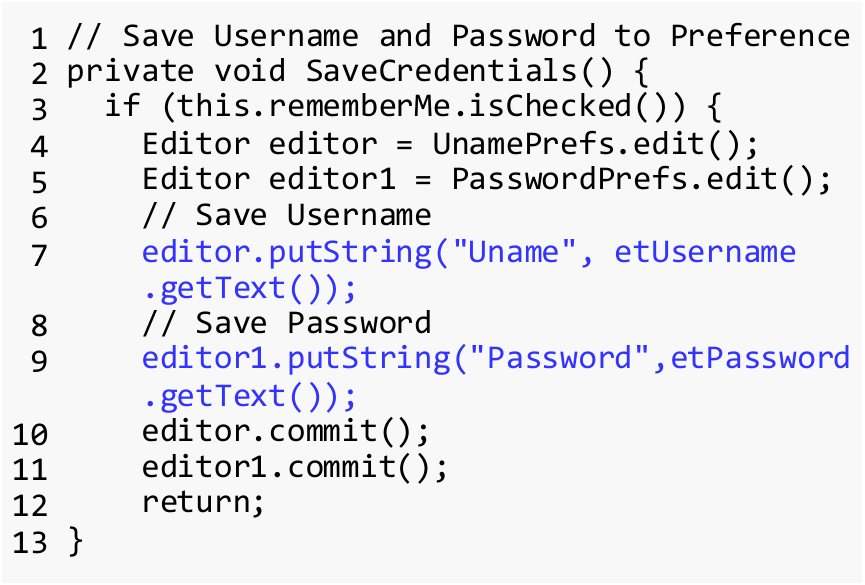}
		\caption{Simplified code of {Preference} weakness in G$\ast$ }\label{fig:example_code_preference}
	\end{minipage}\quad
	\begin{minipage}[t]{.315\linewidth}
		\includegraphics[width=1\linewidth]{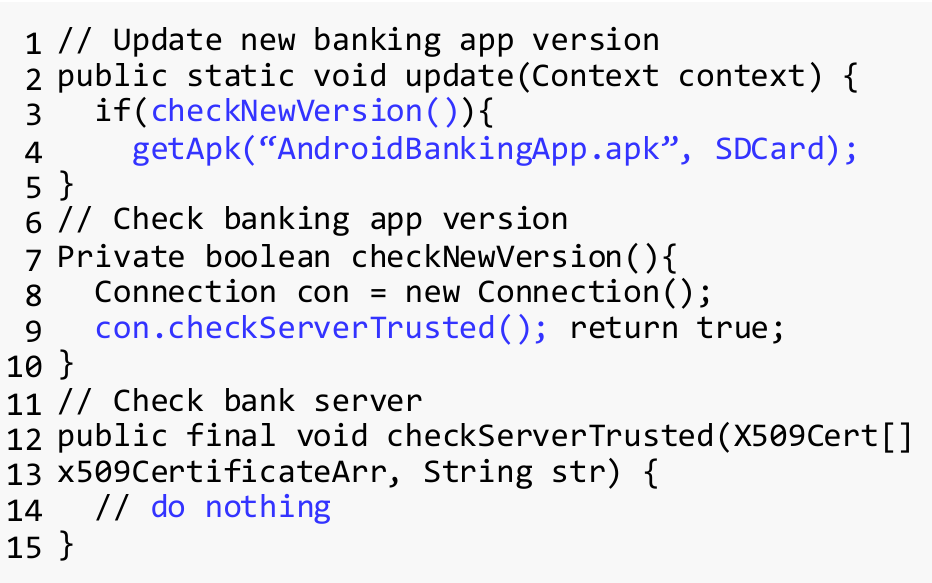}
		\caption{Simplified code of update weakness in I$\ast$ {SMS}}
		\label{fig:example_code_mitm}
	\end{minipage}\quad
	\begin{minipage}[t]{.352\linewidth}
		\includegraphics[width=1\linewidth]{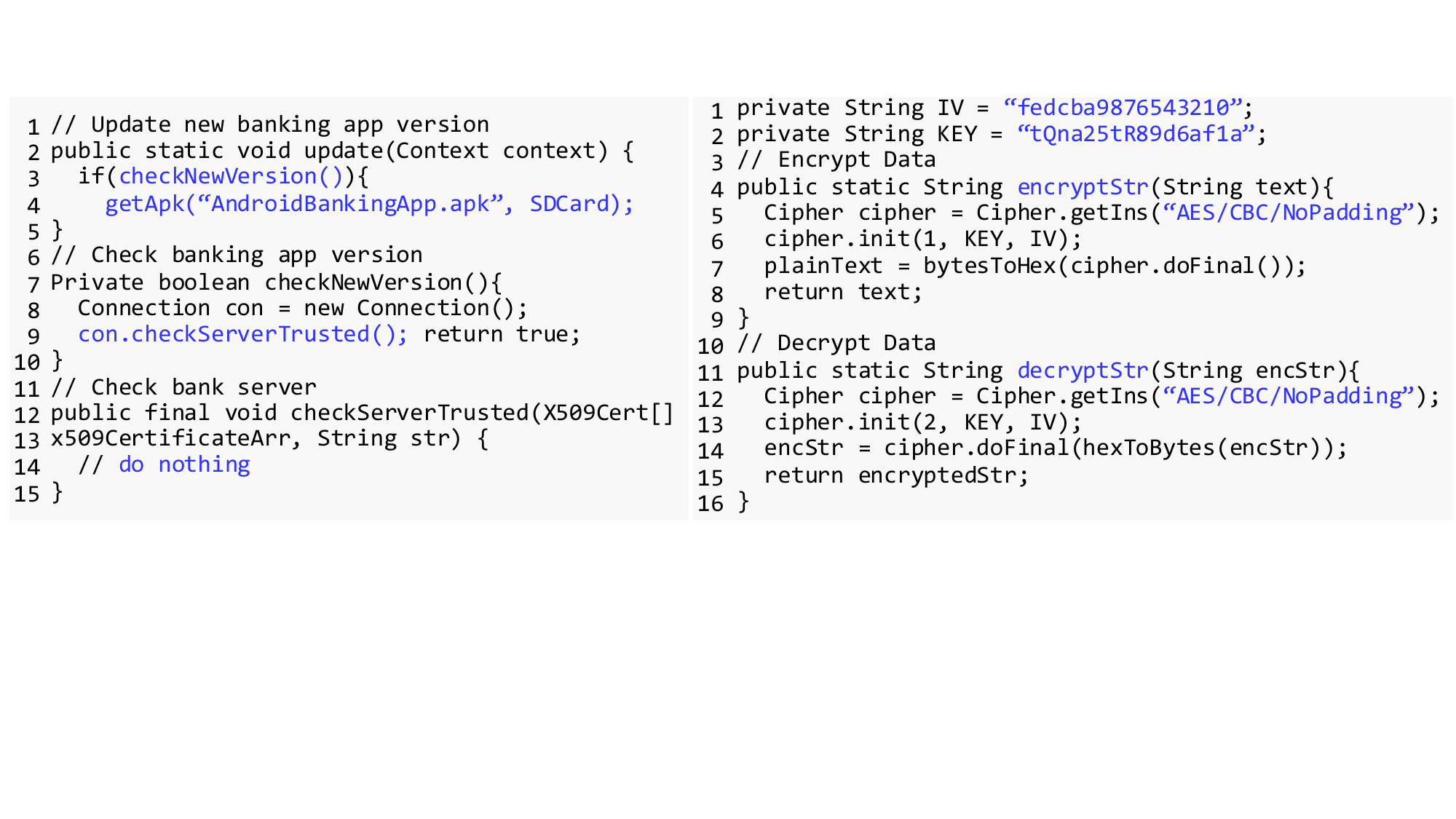}
		\caption{Simplified code of en/decryption weakness in N$\ast$}
		\label{fig:example_code_hard}
	\end{minipage}
\end{figure*}


\subsection{Case Studies of Weaknesses}\label{sec:finding:attack}
To showcase the exploitability of these weaknesses, we introduce \revised{4} vulnerable apps reported by \ausera. 

\noindent\textbf{Screenshot weakness.}
A$\ast$ Bank (v3.3.1.0038) employs two-factor authentication,
\ie, the user first inputs the username and password, and then enters verification code sent by the bank server.
It can be attacked if the login page is not protected (without setting the flag \texttt{ WindowManager.LayoutParams.FLAG\_SECURE} to prohibit taking \emph{screenshot}), and
the verification code can be accessed with granted permissions.
As such, we generate a \revised{malicious app~\cite{chen2019gui, tang2019large}} that runs a service which \revised{can take screenshot of} the screen and read the verification code from SMS during the process of login. As a result, the remote attacker can steal the credentials and bypass the login authentication. {Note that the crafted malware~\cite{chen2016stormdroid, fan2016poster, feng2019mobidroid}} has bypassed the security vetting of Google Play and is successfully put on the shelf, \revised{which makes this attack more practical~\cite{chen2016towards,chen2018automated, chen2019can}.}

\noindent\textbf{Preference weakness.}
Figure~\ref{fig:example_code_preference} shows the vulnerable code of a \texttt{Preference} weakness in G$\ast$ Bank (v1.1) from Algeria. This app stores the credentials (\ie, username and password) into \texttt{Preference} named \texttt{UnamePrefs} and \texttt{PasswordPrefs} (lines 6-9). 
To steal these credentials, we can either (1) create a malicious app signed with the same key, so that it can run in the same sandbox as the victim app on a non-rooted device; or (2) create a malicious app that modifies the original file permission from ``660'' to ``777'' by running \texttt{Runtime.getRuntime().exec} on rooted devices~\cite{reaves2015mo,reaves2017mo}.
In either way, the malware can access the victim's sensitive data stored in the \texttt{Preference}.
Even worse, we find that several apps use insecure permissions \texttt{MODE\_WORLD\_READABLE/WRITEABLE} rather than \texttt{MODE\_PRIVATE}, which 
eases such attacks.

\noindent\textbf{Version update weakness.}
\revised{I$\ast$ SMS Bank (v5.0) is detected as having a MITM risk during version updates, the vulnerable code is shown in Figure~\ref{fig:example_code_mitm}.} 
The app checks new versions with the bank server once started (line 3), but does not verify the \texttt{X.509} certificates from {SSL} servers (lines 11-15). It allows MITM attackers to spoof the server by crafting an arbitrary certificate.
As a result, the new version can be downloaded to SD Card from an attack server (line~4).
To exploit this, we use \textsc{\small Burp Suite}~\cite{burp} and \textsc{\small Fiddler}~\cite{fiddler} to fool the banking app,
by sending a malicious app to impersonate the most recent version~\cite{version}. 
After this malicious app is installed, it serves as a phishing app to steal user credentials and other data.

\noindent\textbf{\revised{Encryption/Decryption} attack.}
\ausera detects an encryption \weakness in N$\ast$ Bank (v1.8) as shown in Figure~\ref*{fig:example_code_hard}. It leaves the hard-coded {AES} keys (\texttt{IV} and \texttt{KEY}) as plain text (lines 1-2), and uses them to encrypt and decrypt
the communication between the app and the bank server.
By leveraging these keys, we successfully decrypt all sensitive data during communication.
Moreover, {AES} uses block cipher modes. If we set with \texttt{NoPadding} (lines~5 and 12), it is easier for attackers to subvert encryption because they only need to decrypt one of the blocks.

\section{{Lessons Learned and Limitations}}

\noindent{\bf \sen{Lessons learned.}}
\sen{(1) According to the security assessment of global banking apps in Table~\ref{tbl:number}, banking apps are not as secure as we expected in the real world. Meanwhile, the results of the global status and longitudinal studies unveil many security threats and unreasonable phenomena. Stockholders such as security teams in banks should pay more attention on these security issues.
	(2) The processes of weakness reporting and patches tracking reveal the gaps between academic researchers, banks, and third-party security companies. 
	(3) The processes of meeting and discussions between corresponding banks bring useful recommendations, and some of them have been used to improve the banking app security. 
	(4) From the perspective of banks, they should pay more attention to security issues compared with functional bugs. Meanwhile, they should provide various channels to respond to the reported vulnerabilities, to make the patching process more efficient. 
	(5) Fortunately, some of banks have accepted our reported vulnerabilities and actively collaborated with us to improve their app security by using \ausera before releasing new app versions.}

\noindent{\bf \sen{Limitations.}}
\sen{(1) The proposed data-related baseline is integrated by many channels based on our depth understanding and knowledge, thus might be incomplete. However, we can investigate the global ecosystem of banking apps based on the baseline. Meanwhile, according to the communications with real banks, they {are highly concerned about} the security weaknesses we proposed in Table~\ref{tbl:issues}.
	(2) The keyword database is constructed first with manual selection of keywords, and then extended with the help of NLP techniques. However, some of keywords may be ignored in the manual analysis process. Actually, the database can be further extended with the increasing banking apps.
	(3) \ausera is built on the top of the static analysis framework (i.e. \textsc{Soot}), thus inherits the limitation of \textsc{Soot} that it may fail and lose some data flows, creating false negatives.}

\section{Related Work}\label{sec:related}

\noindent\textbf{Security assessment of banking apps.} 
In 2015, Reaves et al.~\cite{reaves2015mo} realized the severe \weaknesses of branchless banking apps. 
They reverse engineered and then manually analyzed 7 apps from developing countries, and last found 28 significant \weaknesses. 
Most of these \weaknesses remained unresolved after one year~\cite{reaves2017mo}. Chanajitt et al.~\cite{chanajitt2018} also manually analyzed 7 banking apps, and investigated three types of \weaknesses, including how much sensitive data is stored on device, whether the original apps can be substituted, and whether communication with the remote server can be intercepted. 
Our study differs from \cite{reaves2015mo,reaves2017mo,chanajitt2018} with regards to the scope of the study. Whereas \cite{reaves2015mo,reaves2017mo,chanajitt2018} mainly leverage case studies to study banking apps, the focus of our paper is to conduct a large-scale empirical study on security weaknesses of banking apps. 
Furthermore, we also incorporate multidisciplinary expertise (\eg, code comprehension, regulations, economics) to interpret the potential causes of occurrence of security \weaknesses. 
Our work also differs from alternative topics, such as functional bugs~\cite{Fan2018,fan2018efficiently,su2017guided}, performance~\cite{Liu:ICSE2014} and fragmentation~\cite{Wei16}.
For the concrete security weaknesses, for example, {SSL} issues have been widely discussed in~\cite{fahl2013rethinking}, which suggests {revisiting} the {SSL} handling in applied platforms (\eg, iOS and Android). Followed by recent reports~\cite{parasa2016mobile,Lebeck2015} and our observation, we find that many \apps have fairly weak or even no authentication and encryption mechanisms.
Sounthiraraj et al.~\cite{sounthiraraj2014smv} proposed to combine static and dynamic analysis to identify security problems in {SSL/TLS} for Android apps.
Georgiev et al.~\cite{georgiev2012most} focused on {SSL} connection authentication of non-browser software, indicating that SSL certificate validation is defective and vulnerabilities are logical errors, due to the poor design of APIs to {SSL} libraries and misuse of such APIs.
Egele et al.~\cite{egele2013empirical} checked for violations of 6 cryptographic rules (using cryptographic APIs) in real-world Android apps. They applied static analysis to extract necessary information to evaluate the properties and showed that about 88\% of the apps violate the security rules. For our research, we also integrate these aforementioned \weaknesses as vulnerable security points, and examine whether banking apps contain these vulnerabilities.

\vspace{1mm}
\noindent\textbf{Global analysis of banking apps.}
Castle et al.~\cite{castle2016let} conducted a manual analysis of 197 Android apps and interviewed 7 app developers across developing countries (Africa and South America). 
They divided 13 hypothetical attacks into 5 categories and concluded that realistic concerns are on SMS interceptions, server attacks, MITM attacks, unauthorized access, etc. 
Lebeck et al.~\cite{Lebeck2015} summarized \weaknesses of mobile money apps in developing economies, and combined existing techniques (\eg, cryptocurrencies) to achieve security and functionality goals.
Parasa et al.~\cite{parasa2016mobile} studied 9 mostly-used mobile money apps across 9 Australasian countries, and reported the security \weaknesses in authentication, data integrity, poor protocol implementation, malfunction, and overlooked attack vectors. 
They reported that the apps from comparatively developed countries (\eg, \textsc{AliPay}, \textsc{Osaifu-Keitai}) also have \weaknesses. 
Besides, Taylor et al.~\cite{taylor2017longitudinal} adopted two off-the-shelf tools to roughly scan the apps that are labeled as finance from Google Play Store.
All these prior work adopts small-scale analysis or is taken by survey, while our results are obtained in an automated and largest-scale fashion, which have not been systematically scrutinized before.
\revised{Besides, Chen et al.~\cite{chen2018mobile} focused on studying the details of issue-reporting and issue-patching lifecycle based on the results of weakness detection tools like \ausera~\cite{chen2018ausera}. It unveils gaps between the industry and academia regarding the inconsistent understanding of reported issues and responsibilities. However, in this paper, we propose a comprehensive taxonomy of data-related security weaknesses for banking apps, and propose a detection approach based on the taxonomy. Using \ausera, we conducted experiments to identify security weaknesses and investigate the overall ecosystem of global banking apps from multiple aspects.}

\vspace{1mm}
\noindent\textbf{Security analysis of Android apps.}
Taint analysis is a commonly-used method to reveal potential privacy leakage in Android apps. 
For example, \textsc{TaintDroid}~\cite{enck2014taintdroid} is a dynamic taint-tracing tool which tracks flows of private data by modifying \textsc{Dalvik} virtual machine;
\textsc{FlowDroid} and \textsc{IccTA}~\cite{arzt2014flowdroid,li2015iccta} are both static taint analysis tools that accept the \emph{source} and \emph{sink} configurations for privacy leaks. 
However, these tools target on general apps~\cite{chen2019storydroid}, and thus may not be able to unveil specific security weaknesses (summarized in Table~\ref{tbl:issues}) when applied for banking apps. We also detail the differences in Section~\ref{sec:evaluation}.

\section{Conclusion}\label{sec:conclusion}
In this paper, we conduct a large-scale comprehensive empirical study  on the collected 2,157 security weaknesses of 693 banking apps across more than 80 countries from various aspects. 
To collect the dataset, we also propose a three-phase system, \ausera, to automatically identify data-related weaknesses in banking apps.
Our detected security weaknesses (i.e., 52 security weaknesses) have been confirmed and patched by the 21 corresponding banks and some of them have actively collaborated with us to improve the security of their banking apps. 
The study also narrows down the gaps between academic research and industrial banks, and helps both banks and third-party companies to better tackle security weaknesses.

\clearpage

\begin{acks}
	We appreciate all the reviewers for their valuable comments. We would like to acknowledge HSBC Cybersecurity team for their conscientious response to our responsible disclosure.
	This work is partially supported by the National Satellite of Excellence in Trustworthy Software System (Award No. {NRF2018NCR-NSOE003-0001}), the National Research Foundation, Prime Ministers Office, Singapore under its National Cybersecurity R\&D Program (Award No. {NRF2018NCR-NCR005-0001}) and the Singapore National Research Foundation under NCR Award Number
	{NRF2018NCR-NSOE004-0001}.
\end{acks}
		
		\bibliographystyle{ACM-Reference-Format}
		\balance
		\bibliography{ref}
		
	\end{document}